\documentclass{PoS}
\usepackage[authoryear,square]{natbib}
\bibpunct{(}{)}{;}{a}{}{,}

\usepackage{floatrow}
\DeclareFloatFont{tiny}{\tiny}% "scriptsize" is defined by floatrow, "tiny" not
\title{Observations of the Intergalactic Medium and the Cosmic Web in the SKA era}

\ShortTitle{The Intergalactic Medium}

\author{\speaker{A. Popping}$^{1,2}$, {M. Meyer}$^{1,2}$, {L. Staveley-Smith}$^{1,2}$, {D. Obreschkow}$^{1,2}$, {G. I. J{\'o}zsa}$^{3,4}$, {D.J. Pisano}$^{5,6}$\\%\thanks{A footnote may follow.}\\
      \ \\
       $^1$ICRAR, The University of Western Australia\\
       $^2$ARC Centre of Excellence for All-sky Astrophysics (CAASTRO)\\
       $^3$SKA South Africa\\
       $^4$Rhodes University\\
       $^5$Dept. of Physics \& Astronomy, West Virginia University\\
       $^6$NRAO-Green Bank.\\
       \ \\
        E-mail: \email{attila.popping@icrar.org}}

\abstract{The interaction of galaxies with their environment, the Intergalactic Medium (IGM), is a very important aspect of galaxy formation. One of the most fundamental, but unanswered questions in the evolution of galaxies is how gas circulates in and around galaxies and how it enters the galaxies to support star formation. We have several lines of evidence that the observed evolution of star formation requires gas accretion from the IGM at all times and on all cosmic scales. This gas remains largely unaccounted for and the outstanding questions are where this gas resides and what  the physical mechanisms of accretion are. The gas is expected to be embedded in an extended cosmic web made of sheets and filaments. Such large-scale filaments of gas are expected by cosmological numerical simulations, which have made significant progress in recent years.  Such simulations do not only model the large scale structure of the cosmic web, but also investigate the neutral gas component. To truly make significant progress in understanding the distribution of neutral hydrogen in the IGM, column densities of N$_{HI}\sim$10$^{18}$ cm$^{-2}$ and below have to be probed over large areas on the sky at sub-arcminute resolution. These are the densities of the faintest structures known today around nearby galaxies, though mostly found with single dish telescopes which do not have the resolution to resolve these structures and investigate any kinematics. Existing interferometers lack the collecting power or short baselines to achieve brightness sensitivities typically below N$_{HI} \sim$10$^{19}$ cm$^{-2}$. Reaching lower column densities with current facilities is feasible, however requires prohibitively long observing times. The SKA will for the first time break these barriers, enabling interferometric observations an order of magnitude deeper than current interferometers and with an order of magnitude better linear resolution than single-dish telescopes.}

\FullConference{
Advancing Astrophysics with the Square Kilometre Array\\
June 8-13, 2014\\
Giardini Naxos, Italy}

\newcommand{\skipthis}[1]{}

\begin{document}

\section{Introduction}
The Intergalactic Medium contains a large fraction of the Universal baryon budget and is a key aspect in galaxy formation. This chapter focuses on recent observational results and progress in simulations of the cosmic web. We will motivate the new insights in cosmology and galaxy evolution that the SKA will bring and we will assess the detectability of the IGM at low column densities with SKA-1 and SKA.

Currently almost everything that is known about the structure of the IGM has been derived from the Lyman resonance scattering lines in the optical and ultraviolet (UV).  Although this will continue to be an important method, there will be a significant change in the near future when radio telescopes will be used for the detection of intergalactic gas. New arrays, such as the LOw Frequency ARray (LOFAR), the Murchison Widefield Array (MWA) and the Precision Array to Probe the Epoch of Reionisation (PAPER) will be used in an effort  to detect the 21-cm signature of the IGM at the epoch of reionization (EoR). Current efforts at different wavelenghts (UV, optical and IR) search for the first galaxies at high redshift and are complementary to the EoR detection experiments with the SKA and its precursors. In studies of the evolution of the luminosity function, redshifts in the range of $z=7$ - $8$ are reached, where it is expected that the 21-cm signature of neutral hydrogen will be seen at slightly higher redshift. 

In this chapter we make predictions for observations that can be performed with SKA1-MID and SKA1-SUR at redshift $z\sim0$ and that will directly detect faint neutral hydrogen gas in emission. Simulations expect that intergalactic gas is accreting onto galaxies, one of the 'holy grails' in understanding galaxy formation is to detect this infalling gas. This has been proven to be extremely difficult, partly due to the interface of outflowing gas caused by galactic fountains which are associated with regions of star formation \citep{2006MNRAS.366..449F}.
To fully understand and test our cosmological model a crucial step is to detect the infalling gas between galaxies and measure its physical properties. This gas can be detected in the local Universe by using the 21-cm line of neutral hydrogen. Apart from some tentative detections (e.g. \cite{2010PhDT.......260P, 2013Natur.497..224W}), there has not been clear evidence for the existence of the diffuse gas filaments and infalling gas. New telescopes such as the Australian SKA Pathfinder (ASKAP), the Meer-Karoo Array Telescope (MeerKAT), the Five hundred meter Aperture Spherical Telescope (FAST) and the SKA will be able to probe the neutral fraction and kinematics of the IGM in the Local Universe at significantly lower column densities compared to what has been done before in HI 21-cm emission. 

This low column density gas, residing in filamentary structures, is the reservoir that fuels future star formation, and could provide a direct signature of smooth cold-mode accretion predicted to dominate gas acquisition in star-forming galaxies today (\cite{2005MNRAS.363....2K, 2009Natur.457..451D, 2009MNRAS.395..160K}). Furthermore, the trace neutral fraction in this phase may provide a long-lived fossil record of tidal interactions and feedback processes such as galactic winds and AGN-driven cavities.\\

In the following section of this chapter we will give a short overview of the distribution of low column density HI gas in the cosmic web and around galaxies. In section two and three we discuss the current status of observations and simulations, that probe these low column densities. We have conducted very detailed performance simulations of the SKA baseline design \citep{2013_ska1_bd_rev1}, which will be discussed in section four. Based on the results of these simulations, we make predictions for several science projects that can be performed with SKA1-MID or SKA1-SUR.  In section six we discuss the prospects of SKA1, the development phase of SKA1 and eventually the full SKA. We end with a short conclusion.

\section{Background}
\subsection{Cosmic Web}

Theories of structure formation predict that matter in the IGM and in galaxies is embedded in a "cosmic web" of walls, sheets and filaments. (e.g. \cite{1985ApJ...292..371D,1985Natur.317..595F, 1996ApJ...471..582M}). This web contains large filaments and structures with sizes that range up to several megaparsecs. Galaxies reside in the high density regions of the cosmic web, at the intersection of sheets and filaments. This large scale galaxy structure has been discovered by redshift surveys (e.g. \cite{1986ApJ...302L...1D}), however the gas that is associated with the filaments has been known much longer and has been detected using the absorption lines in spectra of high redshift quasars (\cite{1965ApJ...142.1633G, 1971ApJ...164L..73L}). The spectral density of these Lyman$\alpha$ lines indicate that at redshift $z=3$, nearly all baryons in the Universe were in a cool phase with temperatures of $T\sim 10^4$ K. \citep{1997ApJ...489....7R}. As time evolves this cool gas is being shock heated during the gravitational collapse of dark matter within the filaments, resulting in a Warm Hot Intergalactic Medium (WHIM) (\cite{2007ARA&A..45..221B}). A part of this gas accretes rapidly onto the virialized dark matter halos of galaxies. Numerical simulations of structure formation support the picture of a cosmic web containing baryonic matter that aligns with the underlying dark matter structure. (e.g. \cite{1994ApJ...437L...9C, 2006ApJ...650..560C, 2004ApJ...616..643F}). An example is shown in Fig 1. from \cite{2009A&A...504...15P} where the total hydrogen component (HII + HI) and neutral hydrogen component (HI) of the cosmic web are reconstructed. In the regions of very high over-density, in the centres of galaxies, the neutral gas density is roughly equivalent to the total gas density, indicating that the gas is fully neutral. However at the edges of galaxies and in the filaments the neutral fraction of the gas decreases rapidly resulting in HI column densities well below N$_{HI} \sim 10^{18}$ cm$^{-2}$.

\begin{figure}[htbp]
\begin{center}
\includegraphics[width=0.49\textwidth]{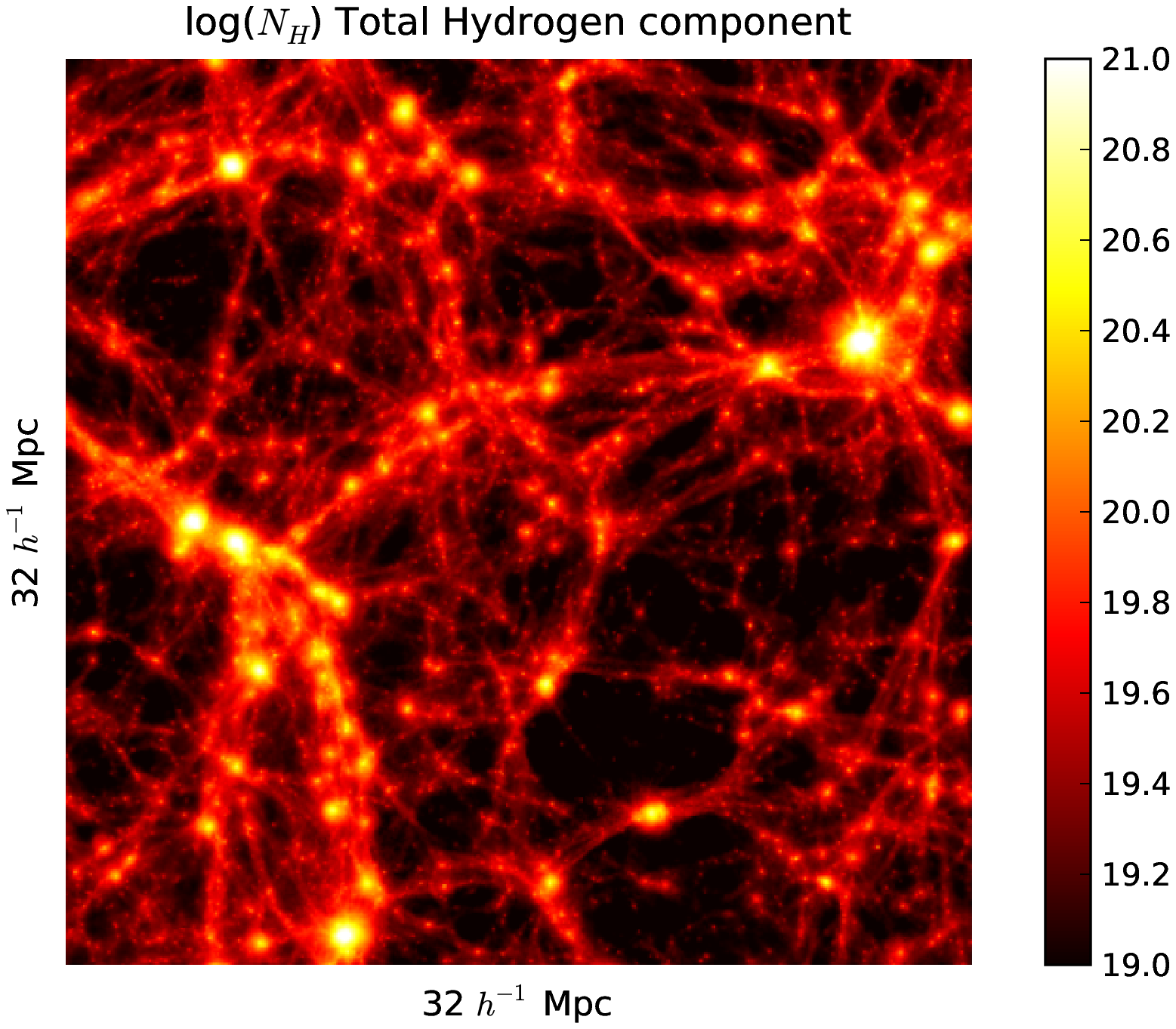}
\includegraphics[width=0.49\textwidth]{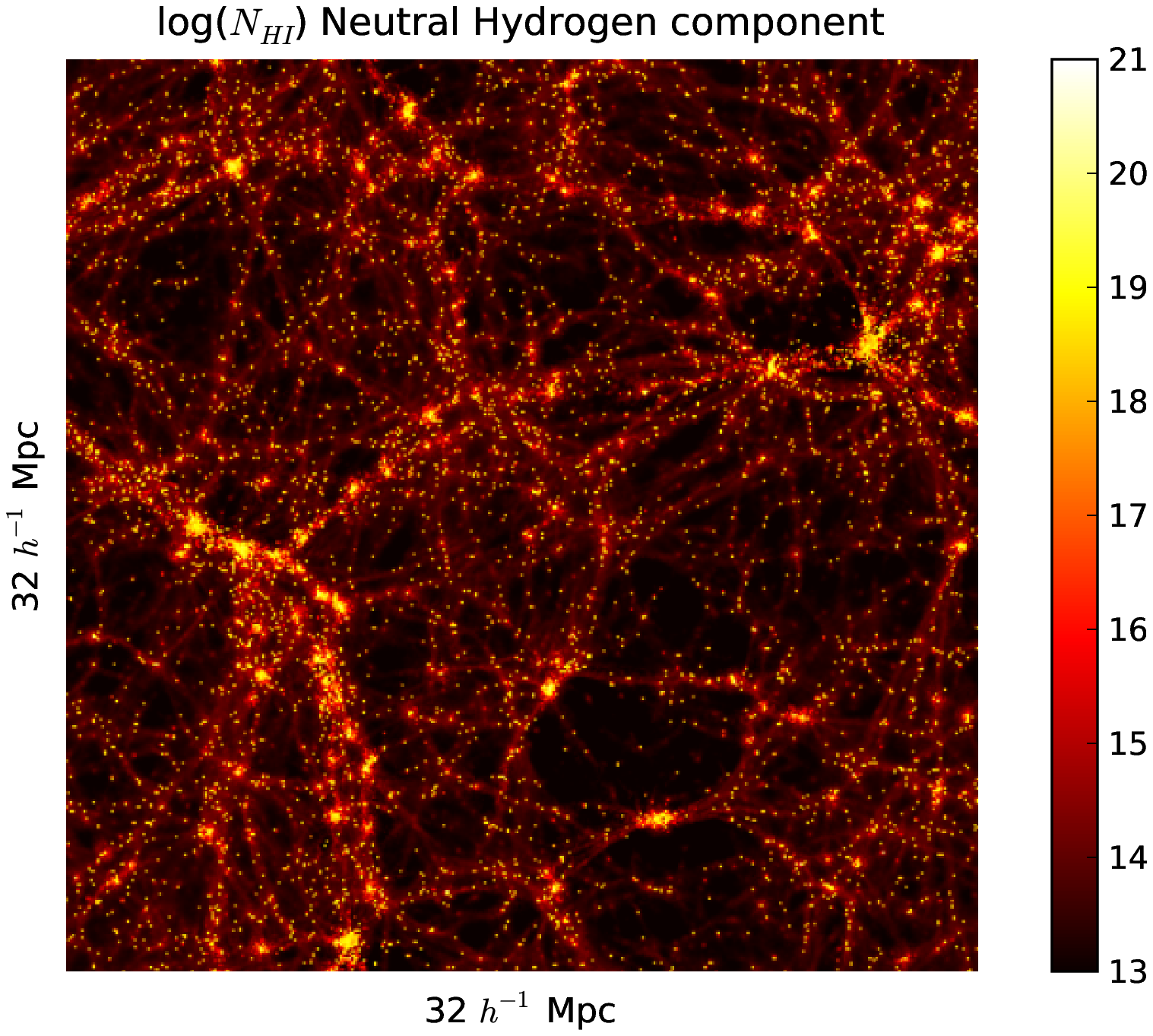}
\caption{Simulation of the cosmic web from \cite{2009A&A...504...15P}, left panel shows column densities of the total Hydrogen gas component (HII + HI), the right panel shows the neutral gas component (HI) that can potentially be detected by radio telescopes using the 21-cm line.}
\label{cosmic_web}
\end{center}
\end{figure}

\subsection{Gas Accretion}
The formation and evolution of galaxies is expected to be intimately connected to the IGM.  Stars in galaxies are formed from molecular gas and after the galaxies have been formed, the stellar mass that is locked in galaxies keeps increasing with time (e.g. \cite{2009ApJ...701.1765M}). This means that to maintain star-formation, the molecular phase needs to be re-supplied over cosmic time. The molecular gas gets replenished from atomic gas, therefore the next question is whether there is enough gas in the dense atomic reservoir directly surrounding the galaxy to explain the continuous star formation. An overview of galaxies with surrounding neutral hydrogen gas is given by \cite{2008A&ARv..15..189S}\\
We have several lines of evidence that the observed evolution of star formation requires gas accretion from the intergalactic medium at all times and on all cosmic scales. Until today, this gas has been unaccounted for and the outstanding question is where this gas resides and what the physical mechanisms of accretion are

When looking at the density of neutral hydrogen 10 billion years ago above redshift $z\sim2$ (e.g. \cite{2009ApJ...696.1543P}), the total amount of gas that is locked in galaxies is less than the mass which is locked in stars today.
Recent simulations support this picture of continuous gas accretion. Due to the gravitational collapse of dark matter dense structures and filaments are created. The primordial atomic gas falls into the gravitational potential wells of the dark matter. There are two modes in which gas falls onto galaxies, dubbed hot-mode and cold-mode accretion \citep{2005MNRAS.363....2K}. In the case of hot mode accretion the gas falling on the dark matter filaments is shock heated to temperatures up to $10^7$ Kelvin and forms a quasi hydrostatic equilibrium halo; the warm-hot intergalactic medium (WHIM) (e.g. \cite{2001ApJ...552..473D}). At some evolutionary stage this hot virialized gas cools rapidly while loosing its pressure support and settles into the centrifugally supported disks or the spiral arms of a galaxy. In the case of cold mode accretion the gas is never heated to these high temperatures, but smoothly accretes directly into the galaxies. The empirical deviation between the two accretion modes is around $2\times10^5$ K.
Recently simulations have converged to a cosmological model where cold mode accretion dominates gas infall at all cosmic times and hot mode accretion is mostly relevant for galaxies that reside in halos with masses above $10^{12}$ solar masses \citep{2009MNRAS.395..160K, 2009Natur.457..451D}.

\subsection{The Circumgalactic Medium}
Surrounding the direct environment of galaxies is the circumgalactic medium (CGM) which is the interplay between galaxies and the IGM. The gaseous haloes of 44 galaxies between $z=0.15$ and 0.35 has been investigated using background QSOs observed with the Cosmic Origins Spectrograph \citep{2013ApJ...777...59T}. The galaxies span both early and late types and the sight-lines of the QSO spectra pass within 150 kpc of the galaxies. The authors find  that the circumgalactic medium exhibits strong HI, with a 100\% covering fraction for star-forming galaxies and 75\% covering for passive galaxies. The kinematics of this gas indicate that it is bound to the host galaxy and the bulk of the detected HI arises in a bound, cool, low-density photoionized diffuse medium. This gas may act as fuel for future star formation \citep{2013ApJS..204...17W} and for star forming galaxies intervening gas is found up to 75\% of the virial radius of the galaxy. The total mass of gas in the CGM is expected to be responsible for at least half of the missing baryons  purported to be missing from dark matter halos at the $10^{12}$ $M_{\odot}$ scale \citep{2014ApJ...792....8W}.

\subsection{Low Column Density of the Intergalactic neutral gas} 

\begin{figure}[htbp]
\begin{center}
\includegraphics[width=0.6\textwidth]{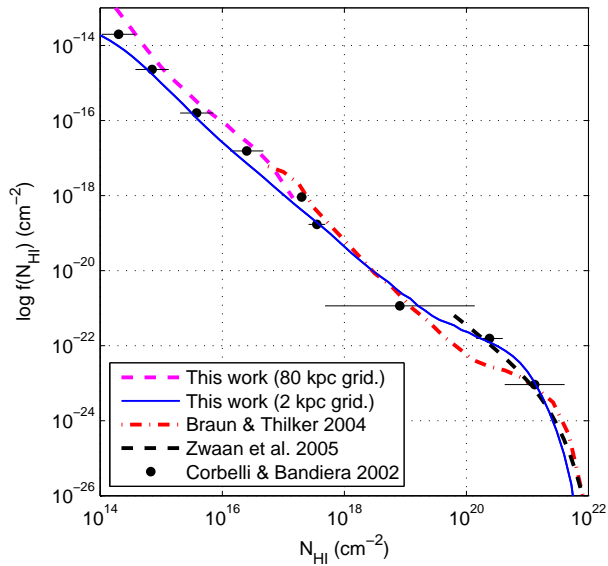}
\caption{Simulated $N_{HI}$ distribution function from \cite{2009A&A...504...15P} at redshift $z=0$. Around column densities of $N_{HI}\sim10^{19}$ cm$^{-2}$ there is a flattening plateau in the distribution, as the gas densities are getting lower as the gas is not protected anymore against ionisation by self shielding.}
\label{NHI_distr}
\end{center}
\end{figure}

The single biggest challenge in detecting neutral hydrogen gas in the IGM is that extremely low brightness sensitivities have to be achieved, well below $10^{19}$ and possibly $10^{18}$ cm$^{-2}$. Unfortunately, emission from intergalactic baryons is difficult to observe, because current interferometers result in a detection limit of column densities typically N$_{HI} \gtrsim 10^{19}$ cm$^{-2}$. These densities are the realm of damped Ly$\alpha$ (DLA) systems and sub-DLAs. Below column densities of N$_{HI} \sim 10^{20}$ cm$^{-2}$, the neutral fraction of hydrogen decreases rapidly because of the transition from optically-thick to optically-thin gas ionised by the metagalactic ultraviolet flux. At lower densities, the gas is no longer affected by self-shielding and the atoms are mostly ionised (e.g. \cite{1998ARA&A..36..267R}). This sharp decline in neutral fraction from almost unity to less than a percent happens within a few kpc. Below N$_{HI} \sim 3\cdot10^{17}$ cm$^{-2}$, the gas is optically thin and the decline in neutral fraction with total column is much more gradual. A consequence of this rapid decline in neutral fraction is a plateau in the HI column density distribution function between N$_{HI} \sim 10^{18}$ and N$_{HI} \sim 10^{20}$ cm$^{-2}$, where the relative surface area at these columns shows only modest growth. This behaviour is confirmed in QSO absorption line studies tabulated by \cite{2002ApJ...567..712C} and in HI emission by \cite{2004A&A...417..421B}. This plateau in the distribution function is a critical issue for observers of neutral hydrogen in emission and the flattening in the distribution function near  N$_{HI} \sim 3\cdot10^{19}$ cm$^{-2}$ has limited the ability of even deeper observations to detect hydrogen emission from a larger area. This behaviour is demonstrated in Fig. \ref{NHI_distr} from \cite{2009A&A...504...15P}, where the distribution on neutral hydrogen gas is simulated. By establishing that a steeper distribution function is again expected below about N$_{HI} \sim 10^{18}$, it provides a clear technical target for what the SKA needs to achieve to effectively probe diffuse gas. Exploration of the N$_{HI} <  10^{18}$ cm$^{-2}$ regime is essential for gaining a deeper understanding of the repository of baryons that drive galaxy formation and evolution.

\section{Current Status of observations and simulations}

\subsection{Observations}

Detecting the gas in the cosmic web in HI and understanding the relation between galaxies and the IGM has proven to be very challenging, because the neutral fraction of this gas is very low. At redshifts above $z\sim2$ HI clouds are visible as intervening structures in Lyman-alpha absorption lines of QSO spectra which are able to probe very low column densities. 
 However these are indirect detections and are confined to a very small number of locations where there is a sufficiently bright background quasar. As a result the Lyman-$\alpha$ lines are generally not suitable to either image the intergalactic gas, or measure any kinematics. Nevertheless, studies with especially the Cosmic Origins Spectograph \citep{2013ApJ...777...59T} have shown very substantial indications for the existence of large amounts of gas surrounding galaxies out to 75\% of the virial radius.\\

 At low redshifts HI observations have been pushing the limits of existing telescopes, to trying to image the extended environment of galaxies at very low column densities and to detect gas filaments connecting the galaxies. The best known example is a filament of HI gas that has been detected between M31 and M33 by \cite{2004A&A...417..421B} using the Westerbork telescope as an array of single dish telescopes rather than an interferometer. The large beam size of the Westerbork telescope when used as a single dish telescope results in a a very good brightness sensitivity, however is limited to very nearby objects such as M31 or the Magellanic clouds. Continuing on this concept \cite{2011A&A...527A..90P} performed a survey of the filament connecting the Local Group with the Virgo cluster, again using the WSRT as a single dish telescope. Some tentative detections have been made for which confirmations were sought in the cross correlation data of the survey \cite{2011A&A...528A..28P} and re-reduced data from the HIPASS survey covering the same area \cite{2011A&A...533A.122P}.
More recently  the connecting bridge between M31 and M33 has been confirmed by \cite{2012AJ....144...52L}, which until today remains the most convincing example of a very extended, low column density HI filament connecting two galaxies. 

Several other attempts have been made to detect gas between galaxies at very low column densities, e.g. \cite{2013Natur.497..224W} and \cite{2014AJ....147...48P}. Both these papers detect some gas features with column densities of $\sim 10^{18}$ cm$^{-2}$. \cite{2013Natur.497..224W} report clumpy structures of gas, however they lack the resolution to resolve most of the kinematics. A very deep observation on Arecibo was conducted by \cite{2009ApJ...692.1447I}, looking for companions of NGC 2903 and reaching column densities of $N_{HI} = 2\cdot 10^{17}$ cm$^{-2}$. The galaxy is found to have a very large HI envelope and two companions that are most likely dwarf galaxies. In a very different approach \cite{2010Natur.466..463C} have attempted to map the cosmic web in HI at redshift $z\sim0.8$ using intensity mapping with the GBT. Noteworthy is that all very low column density detections have been made with single dish telescopes such as the GBT and Arecibo. Existing interferometers lack the sensitivity to detect very low column density gas within a reasonable integration times and have been unable to confirm detections made by single dish telescope. An exception may be a survey of the Virgo area presented in \cite{2011A&A...528A..28P}, where the WSRT has been used at very extreme hour angles. Because of the extreme observing conditions this resulted in a large number of projected short baselines and a low spatial resolution that enabled the detection of low column density gas.\\

Interferometric telescopes have been used to map the extended environment of single galaxies in great detail. Indeed, observations of a few nearby large spirals (e.g. NGC 891 \citep{2007AJ....134.1019O}, NGC 2403 \citep{2004A&A...424..485F, 2014arXiv1407.3648D} and NGC 6946 \citep{2008A&A...490..555B}) and the nearby dwarf galaxy UGCA~105 (e.g. \cite{2014A&A...561A..28S}) have shown large amounts of cold gas in their extended halos. In a review article by \cite{2008A&ARv..15..189S} many examples of galaxies are shown with extended neutral hydrogen emission that are caused by a combination of gas infall and feedback. The authors conclude however that this is not the directly detected infalling gas that fuels the galaxies, because even if all the observed gas would be attributed to gas accretion, an order of magnitude more gas infall is needed to sustain the observed star formation rates.\\

Accretion of intergalactic matter does not only affect the inner portions of galactic disks, but also the outer, extended, HI disks. At large radii (typically beyond $r_{25}$), the spin vectors of outer galactic HI disks often appear to be tilted with respect to the inner disks, i. e.,  galactic disks become warped where the stellar component is fading. The warp, the transition from one orientation to the other, often appears in a confined radial range, indicating a two-regime structure of an inner bright disk and a faint, inclined outer disk (e.g. \cite{1990ApJ...352...15B, 2007A&A...466..883V, 2007A&A...468..903J}). At the transition radius, a sudden drop in surface brightness occurs frequently \citep{2002A&A...394..769G, 2007A&A...466..883V, 2007A&A...468..903J}. These specific features have been interpreted as a signature of accretion from the IGM with misaligned angular momentum \citep{2007A&A...466..883V}.

However, among other scenarios, warps may also be excited through a backreaction of the dark matter halo and the disk disk on accreted dark matter (e.g. \cite{1989MNRAS.237..785O, 2006MNRAS.370....2S}) or by accretion of the IGM onto the - already existing - corona or disk (e.g. \cite{1959ApJ...130..705K}, \cite{2002A&A...386..169L}), which becomes more susceptible to disturbances with increasing radius (see \cite{2007A&A...466..883V}, \cite{2007A&A...468..903J} for more comprehensive reviews). Using the warp structure as a tracer of the IGM is hence an indispensible ingredient in future studies of galaxy evolution, but it requires a good overview of the channels through which galaxies acquire their angular momentum. A more detailed overview about the angular momentum of galaxies is given in \cite{2014AASKA_obreschkow}

Currently the HALOGAS survey \citep{2011A&A...526A.118H} gives the largest inventory of extended gas in nearby galaxies mapping 22 galaxies down to column densities of a few time $10^{19}$ cm$^{-2}$ at a resolution of 15 arcsec. Results of this survey show that an extended HI disk is not common for galaxies (e.g. \cite{2012ApJ...760...37Z}) and that the best example is NGC 891.\\

Apart from targeted observations to achieve low column densities, currently existing all-sky surveys such as HIPASS and ALFALFA also have a good brightness sensitivity. These all-sky surveys have been performed with single dish telescopes, that generally give a much better brightness sensitivity than interferometric observations due to the low resolution of the beam. HIPASS has a typical column density sensitivity of $N_{HI} \sim 4\cdot10^{17}$ cm$^{-2}$ over 26 km s$^{-1}$ \citep{2001MNRAS.322..486B} , this can be slightly improved upon by reducing data in a different way compared to the original product, for example \cite{2011A&A...533A.122P} achieve a brightness sensitivity of $3\cdot10^{17}$ cm$^{-2}$ over 26 km s$^{-1}$ in the northern part of the survey, which is typically less sensitive than the rest of the survey. ALFALFA has a better flux sensitivity compared to HIPASS, but due to the smaller beam it has a very similar brightness sensitivity of $\sim5\cdot10^{17}$ cm$^{-2}$ over 26 km s$^{-1}$ \citep{2005AJ....130.2598G}. Despite the very good brightness sensitivities of these surveys, neither of them show a large amount of column density gas in the realm of a few times $N_{HI} \sim 10^{18}$ cm$^{-2}$ that should be apparent in these surveys. This lack of observational results could support the predicted plateau in the HI distribution function as predicted by simulations. A disadvantege of the surveys performed with single dish telescopes is the very large beam size of up to 15 arcmin the case of HIPASS, that can dilute any structures with physical scales that are smaller than the beam size. The simulations in \cite{2009A&A...504...15P} predict the extended gas clumps or filaments to have widths of the order of $\sim 25$ kpc. At the resolution of the HIPASS this corresponds to physical distance of only 6 Mpc.

\subsection{Predictions from simulations}
In view of the observational difficulties in probing the low HI column density regime, it is particularly important to have reliable numerical simulations to aid in planning new observational campaigns, and eventually to help interpret such observations within a structure formation context. The most widely used methods to make predictions are semi-analytical models and hydrodynamical simulations. In semi-analytical models the evolution of baryons is based on the dark matter distribution of the simulation, hydrodynamical simulations follow the distribution of both baryons and dark matter independently. 
 While simulations of galaxy formation are challenging, historically they have had much success predicting the more diffuse baryons residing in the cosmic web (e.g. \cite{1999ApJ...511..521D}).

The drawback of most numerical simulations is that they concentrate on the total baryon or total gas budget  (e.g. \cite{2009ApJ...698.1467O, 2010MNRAS.406...43P}), but do not focus on the neutral hydrogen component. Only recently the first papers have appeared that make an effort to extract the HI component from hydrodynamical simulations at low redshift (e.g. \citep{2009A&A...504...15P, 2012MNRAS.420.2799D, 2014MNRAS.438.2530C}).  The most significant uncertainty in modelling the neutral hydrogen distribution arises from our need to model a self-shielding correction in moderate density regions \cite{2012MNRAS.420.2799D}. The papers listed here all use a simple prescription to extract the neutral gas component, based on the thermal pressure of the gas. 
Although these predictions have been relatively successful, they do  not justify the true physics of the gas and a lot of effort is going on to model the radiative processes that determine the phase of the gas more realistically.
It also remains a large challenge to understand the interplay between gas in galaxies and the surrounding environment. Although many simulation predict the accretion of gas from the IGM, a significant fraction of the gas in the CGM may be due to feedback mechanisms as a result of supernovae explosion. Very recently \cite{2013MNRAS.432.3005C} have found that in their simulations at $z\sim0$ the electron density of hot coronae around galaxies is dominated by the metal-poor gas accreted from the IGM, they infer that the hot CGM observed via X-ray emission is the outcome of both hierarchical accretion and stellar recycling.

Both \citep{2009A&A...504...15P} and \citep{2012MNRAS.420.2799D} have attempted to model the gas clumps and filaments around galaxies and both papers show small over-densities of HI gas with physical scales in the range between 10 and 50 kpc.
Unfortunately it is very difficult to assess the reliability of the neutral hydrogen distribution in these simulations  as there is a complete lack of observational support at these low column densities. Nevertheless it seems that to align with predictions from simulations, future observations of neutral hydrogen gas in the CGM and IGM should aim to achieve physical resolutions of the order of 10 kpc to be able to resolve the gas. Lower resolution observations might dilute the emission making it harder to detect.

%%%%%%%%%%%%%%%%%%%%%%%%%%%

\section{Performance of the SKA1 baseline design}
In order to assess the potential of phase 1 of the SKA, we have performed detailed simulations to model observations realistically, using different scenarios. For both  SKA1-SUR and SKA1-MID we have simulated observations and compare their performance to that of ASKAP, MeerKAT and the VLA. ASKAP and MeerKAT are the main precursors of SKA1-SUR and SKA1-MID respectively, while the recently upgraded VLA provides the benchmark of what can be achieved with an existing interferometer. 

\subsection{Simulated observations}
 We have created visibility-datasets using the MIRIAD software package, for eight hour synthesis observations with a cycle time of 10 minutes. A continuous integration time of eight hours is considered to be a typical length for observations with SKA1-SUR and SKA1-MID, although shorter integration times are possible and longer integration times can be achieved by combining multiple observations. For each telescope datasets are generated at declinations of 0, -30 and -60 degrees. Beam size, brightness sensitivity and survey speed are dependent on the observing frequency, all simulations have been performed at three different frequencies, relevant to SKA science; 0.6, 1.0 and 1.4 GHz, using a constant bandwidth of 50 kHz. The visibility data is used to generate noise maps and beam-point spread functions (PSF). For all the simulations the telescope parameters have been extracted from table 1,9 and 16 of the SKA1 Baseline Design \citep{2013_ska1_bd_rev1}. For clarification relevant parameters that have been used are listed in table 1. The effective field of view is given in the last row at 1.42 GHz and gives the noise equivalent area of the beam, which can be calculated by:

\begin{equation}
\textrm{FoV}_{beam} = (\pi/8)(1.3 \lambda / D)^2
\end{equation}

\begin{table}[htdp]
\begin{center}
\begin{tabular}{c | l | l | l | l | l }
\hline
\hline
 & {\bf VLA} & {\bf ASKAP} & {\bf MeerKAT} &{\bf  SKA1-SUR} & {\bf SKA1-MID} \\
 \hline
 {\bf A$_{eff}$/T$_{sys}$} & 265 & 65 & 321 & 391 & 1630 \\
 {\bf Receptor Size} & 25 & 12 & 13.5 & 15 & 15 \\
 {\bf T$_{sys}$} & 25 & 50 & 20 & 30 & 20 \\
 {\bf Efficiency} & 0.5 & 0.8 & 0.7 & 0.8 & 0.78 \\
 {\bf FoV$_{beam}$} & 0.16 & 30 & 0.53 & 18 & 0.43 \\
 \hline
 \hline
\end{tabular}
\end{center}
\label{sim_pars}
\caption{Telescope parameters used in the simulations to model observations}
\end{table}%

\begin{figure}[htbp]
\begin{center}
\includegraphics[width=1.0\textwidth]{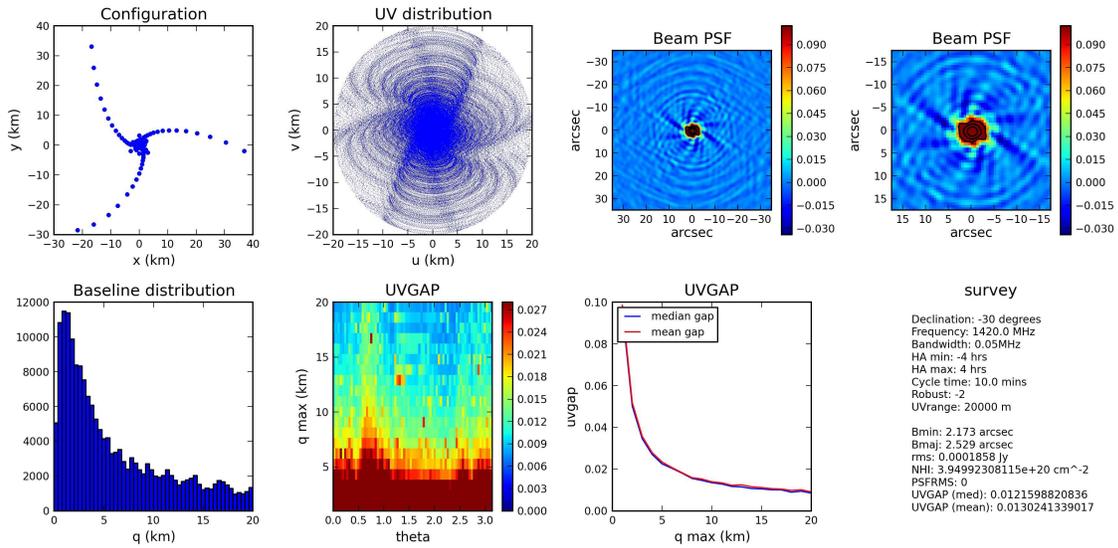}
\caption{Overview panel of the performance results of SKA-SUR for an 8 hour observation at 1.42 GHz using baselines shorter than 20 km. The top panels show the array configuration, the uv distribution and the beam PSF. The bottom panels show a histogram of the baseline distribution, UVGAP values as function of baseline length and polar angle. The third panel at the bottom shows the mean and median uvgap at a certain baseline length. The last panel lists the observation setup and typical numbers that represent the performance of the simulation.}
\label{sur_panel}
\end{center}
\end{figure}

\begin{figure}[htbp]
\begin{center}
\includegraphics[width=1.0\textwidth]{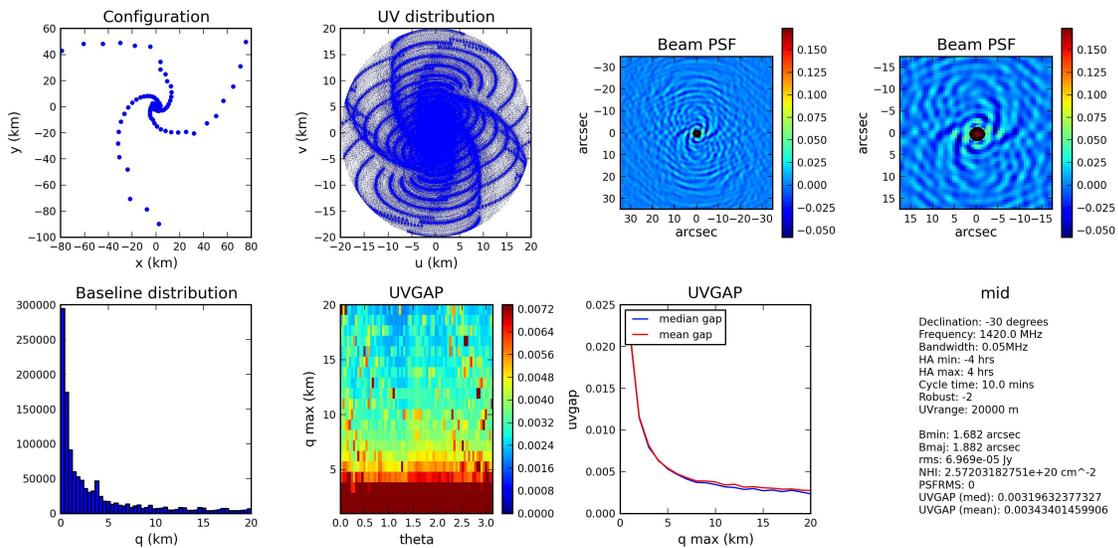}
\caption{Same as figure 2, but now for a simulation representing SKA-MID}
\label{mid_panel}
\end{center}
\end{figure}

To be able to quantify the performance at different resolutions, in the imaging step a tapering function is applied to the visibilities. Both a Gaussian taper and top-hat filter have been applied. The Gaussian taper is the default in MIRIAD, a top-hat filter is a sharp cutoff filter that is applied to the visibility data, to only use $uv$-points within a certain length $q=(u^2+v^2)^{1/2}$. The top-hat filter gives better control over the maximum baseline length and the corresponding resolution when using uniform weighting. We compare the performance of the telescope for a large range of $q$ values, up to baseline lengths of 128 kilometers, or angular resolutions of sub-arcsecond up to several arcminutes.
 Different maps are created using weighting schemes ranging from uniform to natural using 5 robustness parameters (-2,-1,0,1,2), where uniform weighting is equivalent ro robust=-2 and natural weighting corresponds to robust=+2.\\
For every single simulation we have created an overview panel, showing the most relevant information. Two examples are shown in Fig.~\ref{sur_panel} and \ref{mid_panel} for SKA1-SUR and SKA1-MID, where the different panels show different features of the simulated observation, including $uv$-distribution, the beam point spread function and UVGAP. The last panel lists the actual numbers to quantify the performance of a certain telescope configuration and observations setup, listed are:\\
- RMS: the $1\sigma$ noise value in the image maps.\\
- Beam size: the average is taken of the major a minor axis of the synthesized beam\\
- HI column density (N$_{HI}$), the 1$\sigma$ brightness sensitivity is determined at the observing frequency over a 50 kHz bandwidth, taking into account redshift dependence \cite{2013ApJ...770L..29F}:
\begin{equation}
N_{HI} = \frac{2.34 \cdot 10^{20}}{b_{min}b_{maj}}(1+z)^4 S d\nu
\end{equation}
- UVGAP: This parameter gives an impression of how optimal the $uv$ distribution is, it calculates the fraction between the gap between two baselines and the baseline length. UVGAP is a function of baseline-length and polar angle, a full description is given in Millenaar \& Bolton (2010) which is defined as:
\begin{equation}
uvgap = \frac{1}{q_{max}}\sum^n_{k=1} \partial q_k \times (q_k - q_{k-1})\textrm{, with $q_0=0$ and $q_n = q_{max}$}
\end{equation}
The mean and median values of UVGAP as mentioned in the plots are calculated using the top 75 percent of the values in the UVGAP panel (second panel on the bottom), not taking into account the shortest $q$-values.

- Survey Speed: defined here as field of view, divided by the noise squared and scaled with the integration time, (FoV/rms$^2$/$t_{int}$)

\subsection{Performance results}

\begin{figure}[htbp]
\begin{center}
\includegraphics[width=0.66\textwidth]{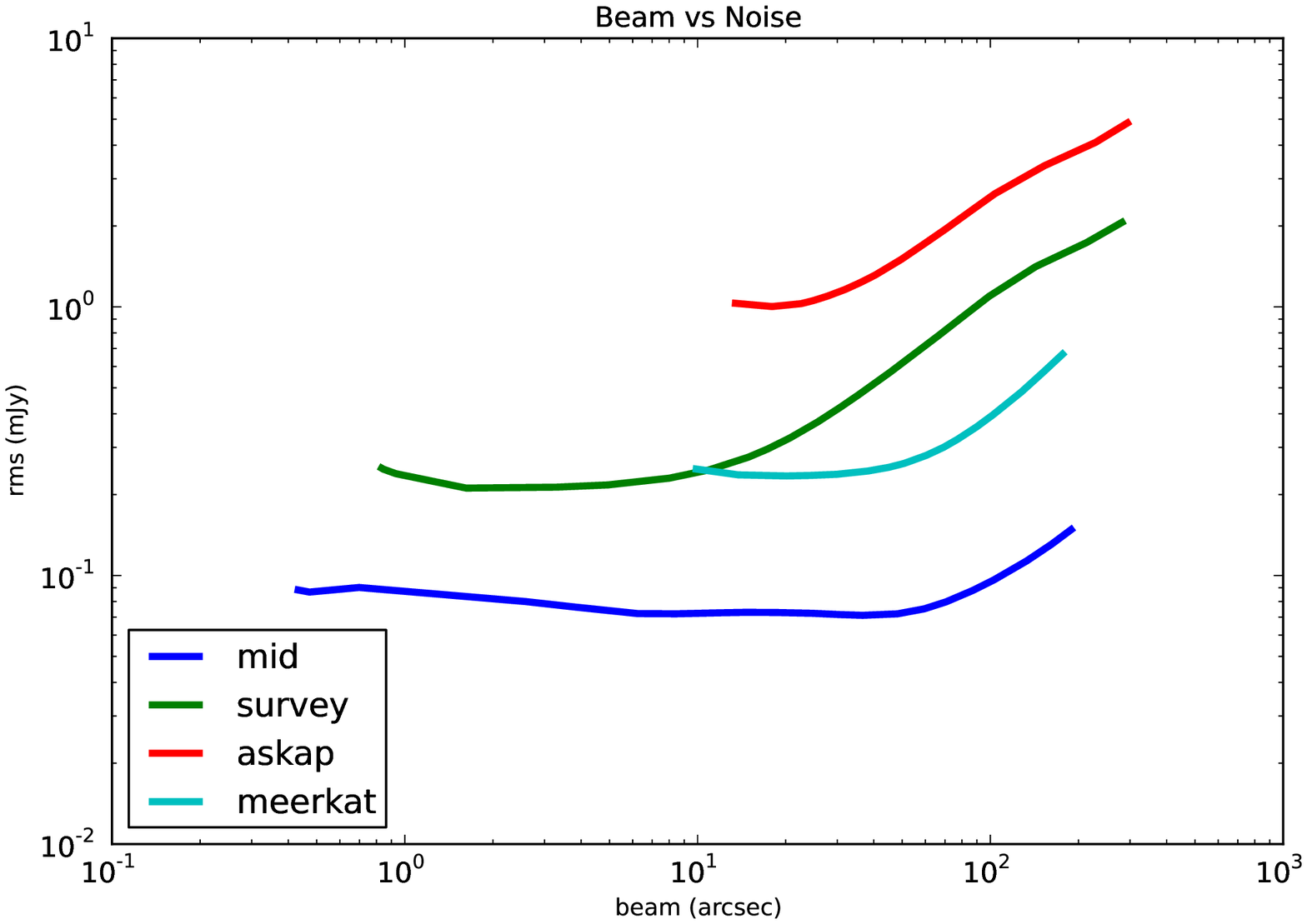}
\includegraphics[width=0.66\textwidth]{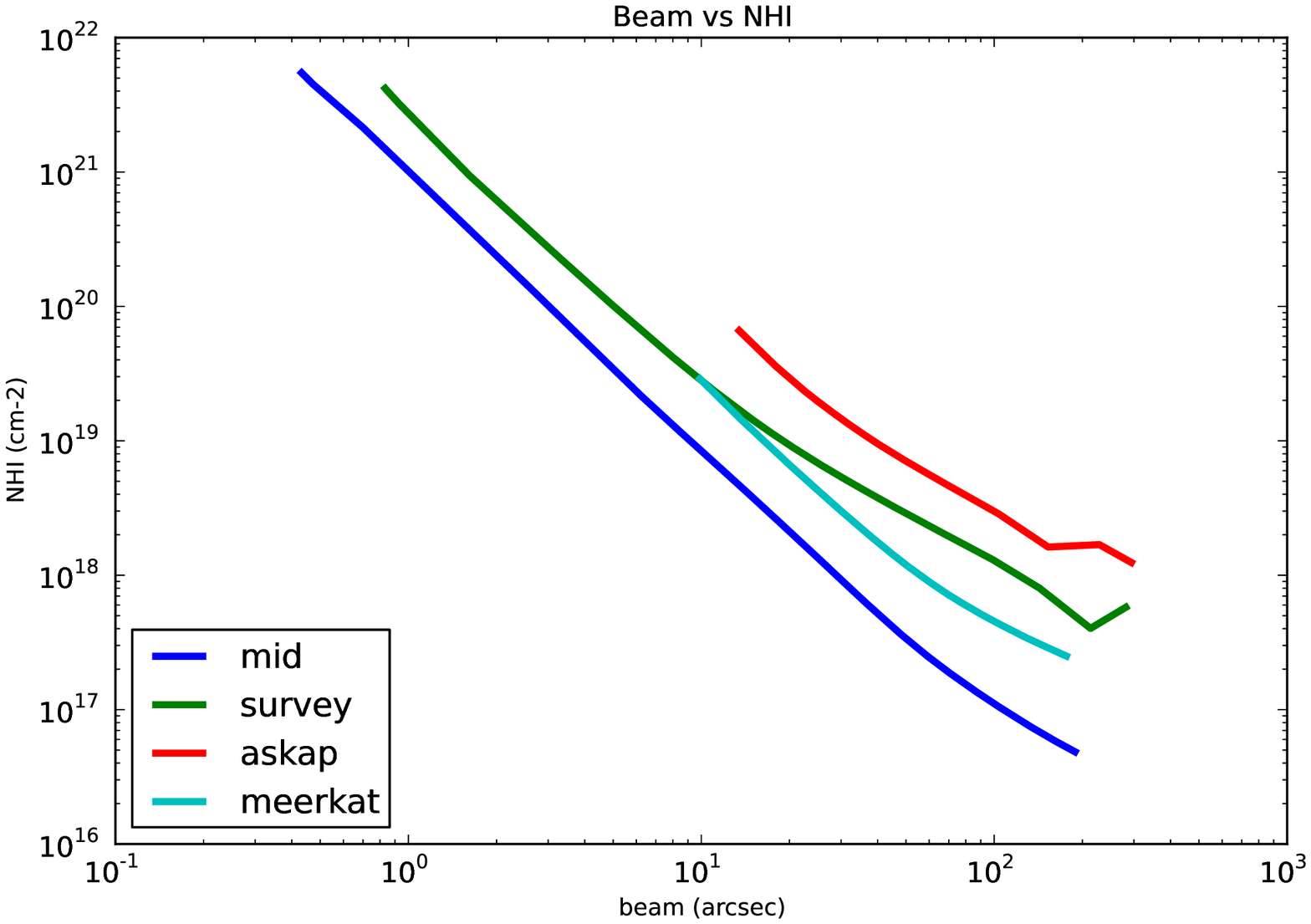}
\includegraphics[width=0.66\textwidth]{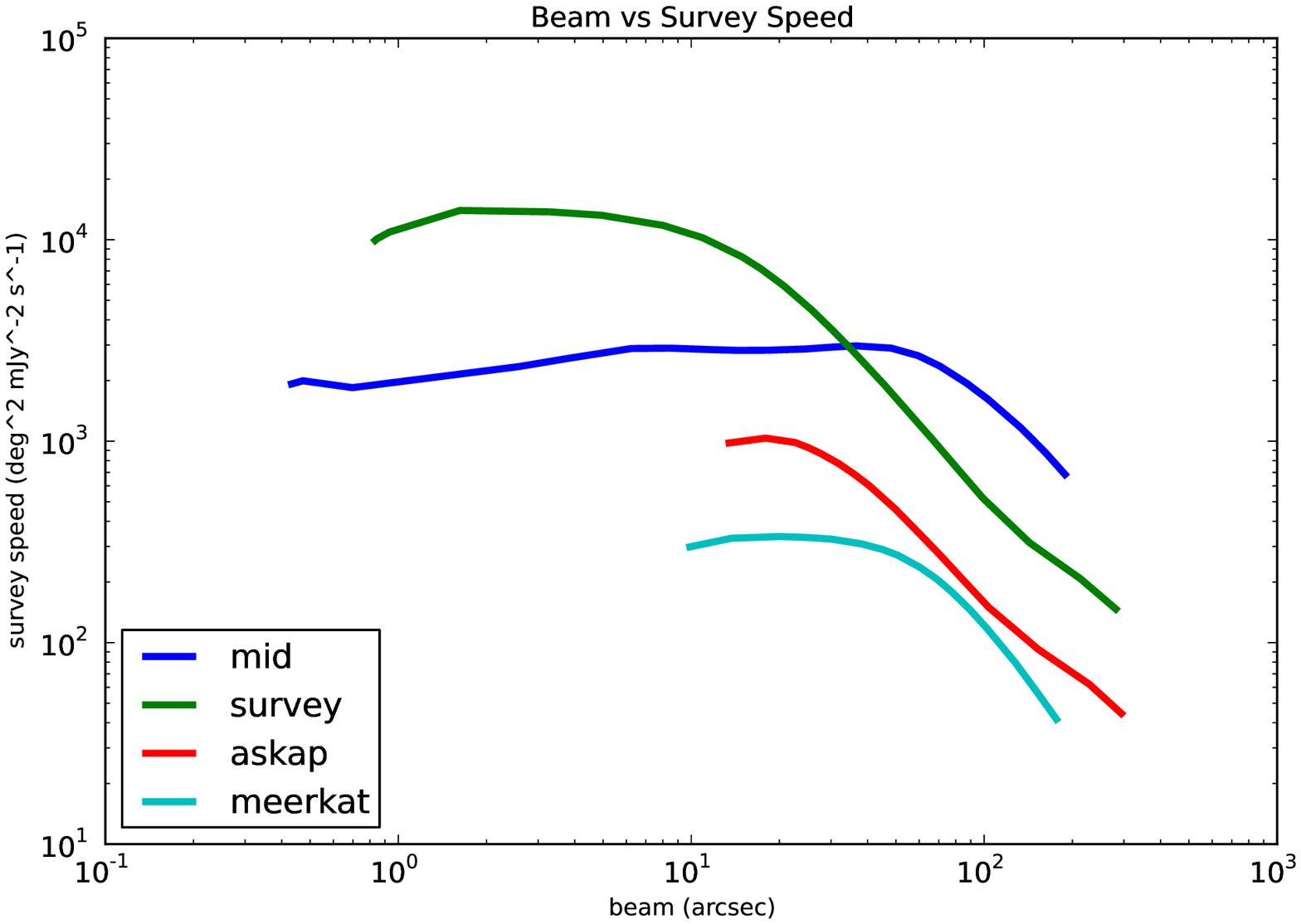}
\caption{Simulated noise, brightness sensitivity and survey speed are plotted as function of angular resolution for different telescopes. All simultaions are performed for an integration time of eight hours, at a declination of -30 degrees and observing frequency of 1.42 GHz, using uniform weighting to generate the image cubes.}
\label{performance}
\end{center}
\end{figure}

Comparing the performance of the SKA to existing telescopes is very relevant to demonstrate where the main differences are and to understand in which parameter space the biggest gain in science can be achieved. In Fig.~\ref{performance} we have plotted the flux rms, brightness sensitivity and survey speed as a function of beam size or resolution for the different telescopes at a declination of -30 degrees, at an observing frequency of 1.42 GHz and using uniform weighting.\\
In the current configuration the flux sensitivity of SKA1-MID remains roughly constant for resolutions smaller than 100 arc sec, while the sensitivity for SKA1-SUR remains stable below $\sim20$ arcsec. This indicates that the distribution of telescopes is close to logarithmic and gives roughly constant flux sensitivity at different scales. The sensitivity of SKA1-MID is about a factor three better compared to SKA1-SUR for the typical resolution range between 1 and 10 arcsecond. This implies, that if the highest possible sensitivity is required, SKA1-MID is the superiour instrument. The same behaviour is reflected in the brightness sensitivity; when using SKA1-MID, a brightness sensitivity of $N_{HI} \sim 10^{19}$ cm$^{-2}$ can be achieved at a resolution of 10 arcsecond after 8 hours of observing.\\ 
The survey speed relates to the inverse of the noise squared, times the field of view, divided by the integration time and shows a very different behaviour. For the nominal observing resolution of a few arcsecond, SKA1-SUR is five or six times faster compared to SKA1-MID. This indicates that when observing at low redshift, any area on the sky larger than three or four degrees squared can be observed faster down to a given sensitivity using SKA1-SUR. This area of three degrees squared is based on a field of view of 18 degrees squared for SKA1-SUR and the factor six difference in survey speed. Obviously this number will change when observing at higher redshift, as the primary beam of SKA1-MID will increase with lower frequency, also increasing the survey speed. For clarification the performance results of SKA1-MID sand SKA1-SUR at 1.42 GHz are listed in table 2.\\

\begin{figure}[htbp]
\begin{center}
\includegraphics[width=1.0\textwidth]{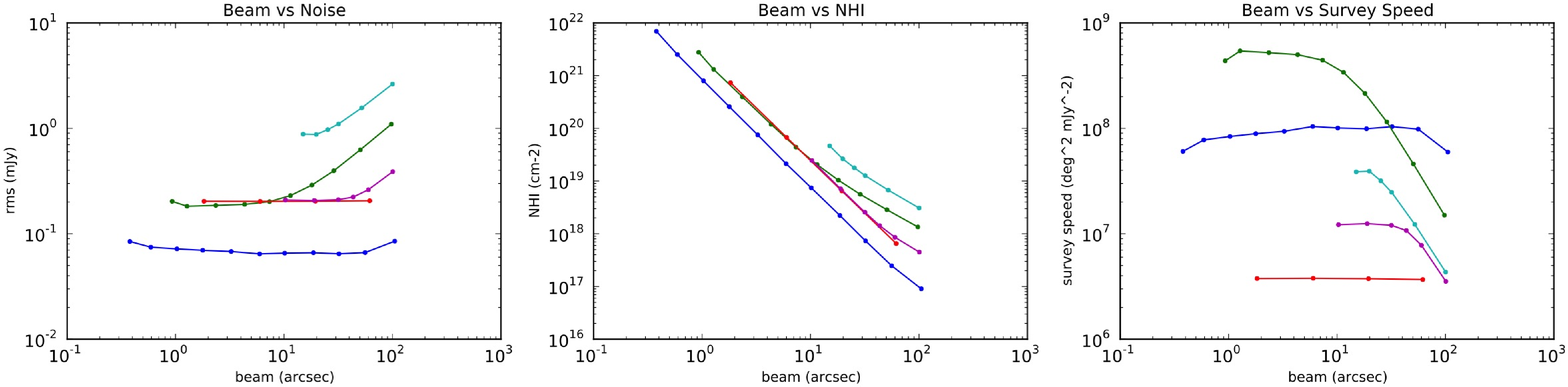}
\includegraphics[width=1.0\textwidth]{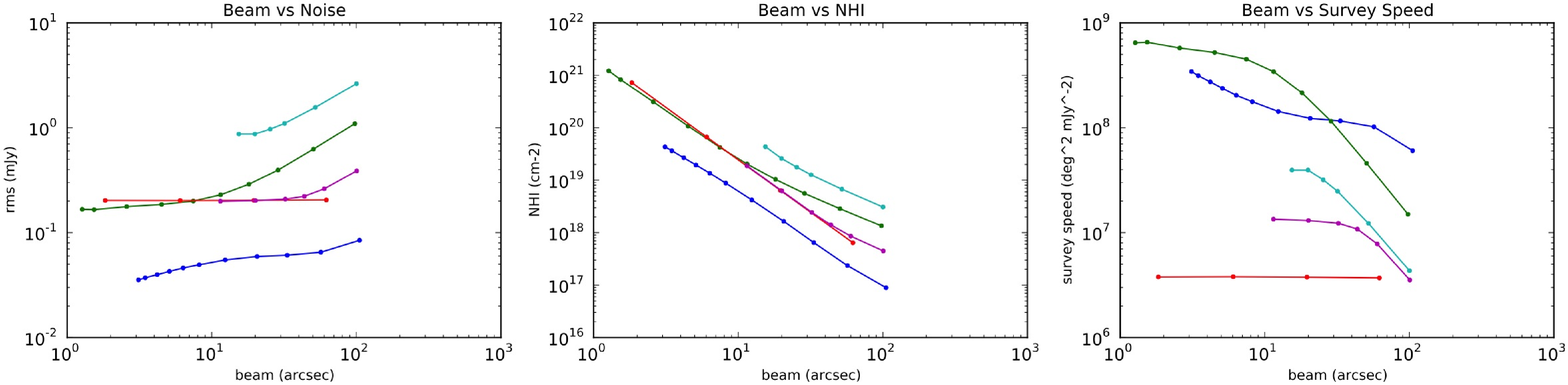}
\includegraphics[width=1.0\textwidth]{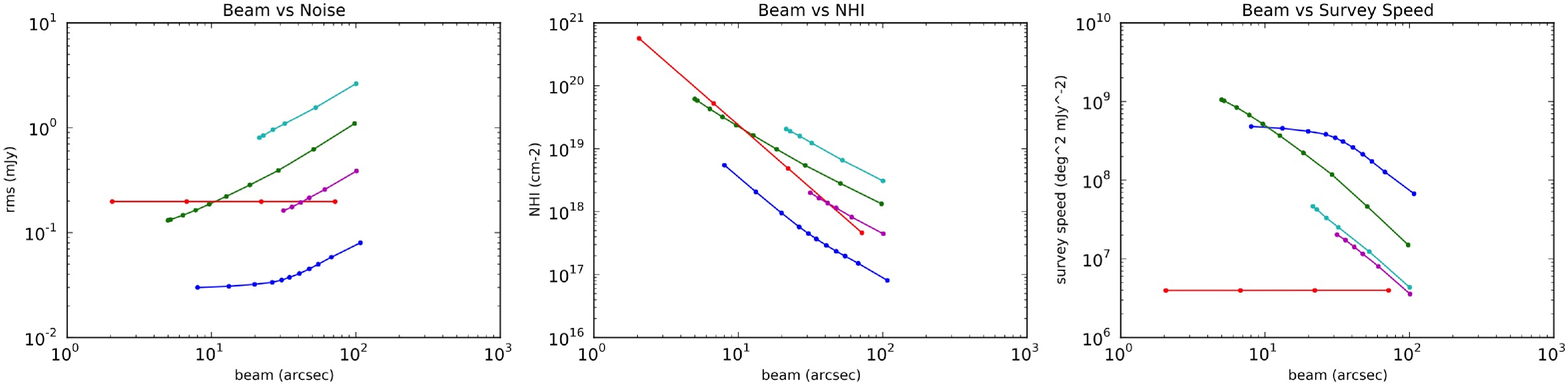}
\caption{Similar as Figure 5, but now different robustness parameters are tested. All simulations are performed at 1.42 GHz, with top panels showing uniform weighting (robust=-2), middle horizontal panels show weighting using robust=0 and the bottom panel represent natural weighting (robust=2). The different lines corresond to SKA-MID (blue), SKA-SUR (green), ASKAP (turquoise) and meerkat (purple). For comparison the perfomance of the JVLA is shown as well (red line).}
\label{weighting}
\end{center}
\end{figure}

In Fig.~\ref{weighting} we have kept all the observing parameters constant, but changed the robustness parameter when making the image cubes using values of -2 (uniform), 0 and 2 (natural). Please note that the x-axis is slightly different in the three rows, because the range in beam sizes is slightly different due to the different weightings. Interestingly the performance of SKA1-MID and SKA1-SUR changes quite significantly at high resolution, but the choice of weighting is not very relevant at low resolution. Low resolution images are mostly affected by the short baselines in the core of the array. Since the core of both SKA1-MID and SKA1-SUR is dense compared to the distribution of antennas in the outer spiral arms, different weighting schemes do not result in significantly different results.

\begin{table}[htdp]
\begin{center}
\begin{tabular}{c c c c}

\hline
\hline
{\bf Performance of SKA1-MID}\\
Beam (arcsec) & rms (mJy) & N$_{HI}$ (cm$^{-2})$  & SS (deg$^2$ mJy$^{-2}$ s$^{-1}$) \\
\hline
0.43    &    0.0883  &   5.47e+21   &  1923.82   \\
0.70    &    0.0902    & 2.16e+21   &  1843.64   \\
1.33    &    0.0850   &  5.59e+20   &  2075.57   \\
2.59    &    0.0799   &  1.39e+20   &  2347.72   \\
3.83    &    0.0762   &  6.04e+19   &  2581.77   \\
6.24    &    0.0721   &  2.15e+19   &  2883.57   \\
8.44    &    0.0720   &  1.18e+19   &  2895.60   \\
12.1    &    0.0726   &  5.82e+18   &  2844.01   \\
14.5    &    0.0729   &  4.01e+18   &  2822.98   \\
18.4    &    0.0728   &  2.52e+18   &  2829.19   \\
24.6    &    0.0723   &  1.39e+18   &  2869.23     \\
30.6    &    0.0715   &  8.87e+17   &  2933.78   \\
36.6    &    0.0711   &  6.18e+17   &  2971.89   \\
48.2    &    0.0720   &  3.62e+17   &  2896.40   \\
59.5    &    0.0751   &  2.47e+17   &  2657.91   \\
70.6    &    0.0798   &  1.87e+17   &  2356.55   \\
87.3    &    0.088     &  1.35e+17   &  1937.84   \\
103.1  &    0.0964   &  1.06e+17   &  1612.83   \\
133.2  &    0.1134    & 7.46e+16   &  1166.96     \\
161.6  &    0.1306   &  5.83e+16   &  879.83   \\
188.7  &   0.1481   & 4.86e+16    &  684.18   \\

\hline
\hline
{\bf Performance of SKA1-SUR}\\
Beam (arcsec) & rms (mJy) & N$_{HI}$ (cm$^{-2})$  & SS (deg$^2$ mJy$^{-2}$ s$^{-1}$) \\
\hline

0.83  & 0.2509   & 4.19e+21  & 9928.39    \\
0.93  & 0.2393   & 3.20e+21  & 10914.27  \\
1.62  & 0.2115   & 9.37e+20  & 13972.02  \\
3.30  & 0.2132   & 2.28e+20  & 13750.09  \\
4.95  & 0.2174   & 1.03e+20  & 13223.94  \\
8.01  & 0.2303   & 4.18e+19  & 11783.98  \\
10.9  & 0.2472   & 2.40e+19  & 10227.82  \\
14.9  & 0.2758   & 1.44e+19  & 8216.59   \\
17.3  & 0.2954   & 1.15e+19  & 7162.42   \\
20.6  & 0.3248   & 8.88e+18  & 5924.45   \\
25.7  & 0.3728   & 6.58e+18  & 4497.05   \\
30.6  & 0.4215   & 5.26e+18  & 3517.91   \\
35.5  & 0.471     & 4.36e+18  & 2817.33   \\ 
45.6  & 0.5723   & 3.22e+18  & 1908.24   \\
55.9  & 0.6773   & 2.53e+18  & 1362.44   \\
66.7  & 0.7838   & 2.06e+18  & 1017.35   \\
82.7  & 0.9406   & 1.61e+18  & 706.43    \\
99.0  & 1.095     & 1.30e+18  & 521.26    \\
142.8 &  1.412   & 8.01e+17  & 313.48    \\
213.3  &  1.736  & 4.02e+17  & 207.39    \\
%281.4  & 2.063   & 5.77e+17  & 146.85    \\

\hline
\end{tabular}
\end{center}
\caption{Performance of SKA1-MID and SKA1-SUR at an declination of -30 degrees and observing frequency of 1.42 GHz for an 8 hour observation with a simulated cycle time of 10 minutes and using uniform weighting to generate the image cubes.}

\label{model_results}
\end{table}%

%%%%%%%%%%%%%%%%%%%%%%%%%%

\section{Science projects}
Image maps that are created using uniform weighting have the best predictable behaviour and as can be seen in the previous section the flux sensitivity remains relatively constant for SKA1-MID over the full resolution range and for SKA1-SUR for beam sizes below $\sim20$ arcsec. The same is true for the survey speed of the telescopes.\\
A great benefit of the current configurations is that no special observations are required with the aim of achieving a good brightness sensitivity. Basically any spectral line observation can be used and the choice of tapering and weighting when making the image cubes will determine the resolution.\\

\subsection{Imaging the extended environment of galaxies}
SKA1-MID has a large number of antennas and an effective observing area that is significantly larger than any existing interferometer. As a result it can achieve brightness sensitivities over small areas that are orders of magnitude better than before and really explore new territory.
The HALOGAS survey observed individual galaxies for 10x12 hours with the WSRT. Characteristic one sigma column densities of this survey ar  $N_{HI} \sim 10^{19}$ cm$^{-2}$ at a resolution of 15''. We will use this number as a benchmark of what is possible in large projects and propose to observe selected nearby galaxies for a period of up to 100 hours with SKA1-MID. Such  observations will achieve brightness sensitivities of $N_{HI} \sim 7 \cdot 10^{16}$ cm$^{-2}$ over 10 km s$^{-1}$ at a resolution of 1 arcmin, or $N_{HI}  \sim 5 \cdot 10^{17}$ cm$^{-2}$ over 10 km s$^{-1}$ at a resolution of 20 arcsec. Compared to existing deep observations on single dish telescopes such as the GBT, these observations have a roughly similar brightness sensitivity, however over a beam area that is about a factor 1000 smaller. Not only should such observations confirm the existing detections from single dish telescopes, but also resolve these structures kinematically. Compared to a project such as HALOGAS, these observations have roughly the same resolution, however with a sensitivity that is more than ten times deeper and over an area that is larger due to the larger primary beam size of SKA compared to the WSRT. Both in sensitivity and resolution these observations will open up complete new parameter space. They will provide unprecedented sensitivity to investigate gas accretion and will be able to resolve the clumps or substructure in the extended environment of galaxies.

\subsection{All sky survey of the IGM}

Given that SKA1-SUR is a survey instrument, we assume that this telescope will be used to perform a large area survey on the sky, possibly an all sky survey. Typical integration times for an all sky survey would be 10 hours per pointing which is sufficient to get the full $uv$-coverage and this number is also very similar to the integration time of the anticipated WALLABY survey on ASKAP. Doing such a survey, a brightness sensitivity can be achieved of $N_{HI} \sim 10^{19}$ cm$^{-2}$ over 10 km s$^{-1}$ at a resolution of 15-20 arc sec. This is almost a factor two improvement in brightness sensitivity over WALLABY at a linear resolution that is two times smaller, or four times smaller in beam area. These resolutions correspond to a physical beam size of about 10 kpc at a luminosity distance of 130 Mpc and therefore should be able to detect and possibly resolve faint clumps or filaments up to these distances.\\ 
The brightness sensitivity of such a survey is very comparable to the sensitivity of the HALOGAS, which is currently the deepest survey of individual galaxies limited to a sample of 22 galaxies. A large area survey with SKA1-SUR will provide extended image maps for not only all galaxies in the local universe but also for all positions across the entire sky. Such a survey will provide unprecedented information of the distribution of faint gas at very high resolution and can shed a light on the different modes of gas accretion. Another major difference compared to HALOGAS is that such a survey will map the environment of galaxies out to the virial radius where gas is still expected to reside.  Since all galaxies in the Local Universe will be mapped, these observations will also demonstrate whether galaxies with halo masses below $M_{sun} \sim 10^{12}$ have more extended gas and are more prone to cold gas accretion as is predicted by the simulations.\\

As mentioned in section 3, studies of the HI structure of the outer warped disks of galaxies are a major ingredient in investigating the interplay of the IGM and the ISM, although the theory has to be understood well. An all-sky HI survey with SKA-survey down to a sensitivity where warps are easily detected [see e.g. \cite{2002A&A...394..769G}, \cite{2007A&A...468..903J}] makes comprehensive warp statistics, studies of the structure of the IGM in different environments \citep{2014MNRAS.440L..21H}, as well as studies of large-scale alignment of outer disk material, possible. Tracing the IGM through spin alignment of outer disks has been attempted using optical observations \citep{1990A&A...236....1B,2008A&A...488..511L}. The wide-field access to the much better suited tracer HI, which is often detected far beyond the optical disk and including the kinematical information, will, however, mean an observational quantum step for studies of the ISM-IGM interface.

The mechanism of IGM accretion onto the disc produces warps \citep{2008A&A...488..511L}. The ubiquity of the warp phenomenon might be used to reconstruct both the IGM density profile and the individual member orbits within galaxy groups \citep{2014MNRAS.440L..21H}. Several groups have sought systematically aligned warps within galaxy samples as a probe of large scale accretion signatures.  When probing column densities of $N_{HI} \sim 10^{19}$ cm$^{-2}$ at a resolution of $\sim$15 arcsec, we can detect and resolve the disks and warps of a large number of galaxies.

\section{SKA prospects}

\subsection{SKA Phase 1}

Phase 1 of the SKA can make a major step in helping us understand the distribution of very low column density gas around and between galaxies as outlined in the previous section. We have modelled the performance of SKA1-SUR and SKA1-MID in very high detail. Compared to existing observations we can achieve orders of magnitude improvement in either resolution or sensitivity which will enable us to explore a new area in parameter space.

\subsection{SKA Early science operations}
Development of the SKA is a long process and final completion of even Phase 1 will be preceded by a period of commissioning, testing and early science operations. Nevertheless studies of the IGM and the cosmic web are an excellent test case for the SKA as the technical requirements are relatively modest.  

Most likely the inner core of the SKA will be constructed first, representing the majority of the short baselines. To detect very faint and diffuse gas, the shorter baselines in interferometric arrays are most relevant. Also computing requirements are relatively modest as low resolution cubes are required to study the faint IGM. These cubes are much easier to process than science projects that rely on high resolution and the longest baselines in the array.  Because of these arguments we anticipate that when  50\% of SKA1 is constructed, we will be able to achieve more than 50\% of the science performance, relevant to low resolution images.

When assuming an early science phase with 50\% of the final sensitivity of Phase 1, still very relevant science can be done when 50\% percent of the brightness sensitivity is achieved.  When using the numbers as calculated in the previous section, for SKA1-Mid we predict a 5$\sigma$ brightness sensitivity of $N_{HI} \sim 2\cdot 10^{18}$ cm$^{-2}$ over 10 km/s at 15'' resolution after 100 hours of integration time over 0.46 deg$^2$ at 1.42 GHz. For SKA Survey we predict $N_{HI} \sim 10^{19}$ cm$^{-2}$ over 10 km/s at 15'' resolution after 10 hours of integration over an area of 18 deg$^2$ at the full frequency range of the telescope. The integration times of 100 and 10 hours for respectively SKA1-Mid and SKA1-SUR are chosen in the same way as for the science project of SKA 1 but can easily be scaled to any other integration time. The estimated sensitivities are still a very major improvement compared to all currently available data.

\subsection{Full SKA}
Phase 1 of the SKA will improve upon existing observations by several orders of magnitude. The low column density gas that can be detected is predicted by simulations but apart from that largely unknown. Completion of Phase 2 of the SKA will improve the sensitivities by another order of magnitude. Assuming similar integration times as in previous sections, an all sky survey will be able to achieve a  brightness sensitivity of  N$_{HI} \sim 10^{17}$ cm$^{-2}$ over 10 km s$^{-1}$ at a resolution of 10 arc sec. Targeted observation of single galaxies will be able to detect HI emission down to column densities of $N_{HI} \sim  10^{16}$ cm$^{-2}$ at 1 arcmin resolution. 

%%%%%%%%%%%%%%%%%%%%%%%%

\section{Conclusion}
The interplay of gas between galaxies and the surrounding IGM is a key aspect in galaxy formation. The cosmic web holds the large reservoir of gas that eventually accretes into galaxies and supports the formation of stars. For the imaging of the diffuse gas filaments in the cosmic web, HI column densities are required in the range $N_{HI} \sim 10^{15} - 10^{18}$ cm$^{-2}$ which is mostly the realm of the full SKA. Nevertheless already Phase 1 of the SKA can provide much better understanding about the IGM. Observations with the SKA1 will not only be able to detect low column density gas over very large areas much deeper than ever before, for the first time they will be able to resolve this gas at sub-arcminute resolution.\\
 Studies of the IGM do not require additional observations and can completely rely on observations performed for other HI projects. Due to the close to logarithmic distribution of telescopes in the current baseline design, the noise remains relatively constant as function of resolution. For any existing observations we can make cubes at a lower resolution to enhance the brightness sensitivity.\\
Using SKA1-MID, brightness sensitivities can be reached that are 10 times better compared to current observations, and this can be achieved within a realistic integration time of 100 hours. These integration times align very well with the integration times that are needed to image the ISM in individual galaxies at high spatial resolution (see \cite{2014AASKA_deblok}). Targeted observations of individual galaxies will improve our sensitivity limits and already with SKA1 column densities of $N_{HI} \sim 10^{17}$ cm$^{-2}$ can be detected. Such observations will image and resolve the clumps and substructure in the environment of galaxies that are predicted by simulations, and these observations might start to probe the cosmic web filaments in which galaxies reside.\\
Using SKA1-SUR with an integration time of ten hours per pointing, brightness sensitivities can be achieved equivalent to the best datasets we currently have, however for very large areas on the sky, rather than a small sample of individual galaxies. The SKA will for the first time give column densities of $N_{HI} \sim 10^{19}$ cm$^{-2}$ at an angular scale of 15'' over very large areas of the sky. This will not only image the extended environment of galaxies to study disks and gas accretion, but will also provide an unbiased search for diffuse intergalactic gas.

\bibliographystyle{apj}
\bibliography{ska_chapter}
%\bibliography{names,ska_chapter}

%\begin{thebibliography}{99}
%
%\bibitem[Carilli \& Rawlings(2004)]{2004NewAR..48..979C} Carilli, C.~L., \& Rawlings, S.\
%2004, \nar, 48, 979 
%
%%\bibitem[]{2012arXiv1212.3497E} Ekers, R.\ 2012, arXiv:1212.3497
%
%\bibitem[]{2012arXiv1212.3497E} Ekers, R.\ 2012, \apj
%
%\end{thebibliography}

\end{document}